\documentclass[12pt]{article}
\usepackage{graphicx}
\usepackage{epstopdf}
\usepackage{amsmath,amssymb,amsfonts,graphicx}

\newcommand{\bnabla}{\mbox{\boldmath $\nabla$}}

\begin{document}

\thispagestyle{empty}
\renewcommand{\refname}{References}

\title{\bf The Casimir effect with quantized charged scalar matter in \\
background magnetic field}


\author{Yu. A. Sitenko and S. A. Yushchenko}

\date{}

\maketitle
\begin{center}
Bogolyubov Institute for Theoretical Physics,\\  National Academy of Sciences of Ukraine,\\ 14-b Metrologichna Str., Kyiv, 03680, Ukraine
\end{center}


\begin{abstract}
We study the influence of a background uniform magnetic field and
boundary conditions on the vacuum of a quantized charged massive
scalar matter field confined between two parallel plates; the
magnetic field is directed orthogonally to the plates. The
admissible set of boundary conditions at the plates is determined by
the requirement that the operator of one-particle energy squared be
self-adjoint and positive definite. We
show that, in the case of a weak magnetic field and a small
separation of the plates, the Casimir force is either attractive or
repulsive, depending on the choice of a boundary condition. In the
case of a strong magnetic field and a large separation of the plates,
the Casimir force is repulsive, being independent of the choice of a
boundary condition, as well as of the distance between the plates.
\end{abstract}


\bigskip

\begin{center}
\end{center}

\bigskip
\medskip

\section{Introduction}

   The Casimir effect \cite{Cas1,Cas2} is a macroscopic effect of quantum field theory,
which is caused by the polarization of the vacuum owing to the
presence of material boundaries of a quantization volume, see review
in Ref.\cite{Bor}. The effect has been confirmed experimentally
with a sufficient precision, see, e.g. Refs.\cite{Bre,Dec}, as well as
other publications cited in Ref.\cite{Bor}, and this opens a way for
various applications in modern nanotechnology.

   Generically, the interest was focused on the Casimir effect with
quantized electromagnetic field, whereas the Casimir effect with
other (nonelectromagnetic) quantized fields was mostly regarded as
an academic exercise that could hardly be validated in laboratory.
However, the nonelectromagnetic fields can be charged, and this
opens a new prospect allowing one to consider the Casimir effect as
that caused by the polarization of the vacuum of quantized charged
matter owing to the presence of both the material boundaries and a
background electromagnetic field inside the quantization volume.

   In this respect it should be recalled that the effect of the
background uniform electromagnetic field alone on the vacuum of quantized
charged matter was studied long ago, see
Refs.\cite{Hei1,Eul,Hei2,Wei,Schw} and Refs.\cite{Dit,Du} for reviews. The case
of a background field filling the whole (infinite) space is hard
to be regarded as realistic, whereas the case of a background field
confined to the bounded quantization volume for charged matter looks
much more reasonable and even can be regarded as realizable in
laboratory. Moreover, there is no way to detect the energy density
which is induced in the vacuum in the first case, whereas in the
second case it results in the pressure from the vacuum on the
boundary, and the latter is in principle detectable. One may suggest
intuitively that the pressure, at least in certain circumstances, is
positive, i.e. directed from the inside to the outside of the
quantization volume. A natural question is, whether the pressure
depends on a boundary condition imposed on the quantized charged
matter field at the boundary?

   Thus, an issue of a choice of boundary conditions acquires a primary
importance, requiring a thorough examination. To deal with this
issue comprehensively, one has to care for the self-adjointness of
a differential operator corresponding to the physical observable of a
quantum system. A quest for the self-adjointness
is stipulated by the mere fact that a multiple action is
well-defined for the self-adjoint operator only, allowing for the
construction of functions of the operator, such as evolution,
zeta-function and heat kernel operators, etc. A relevant basic
operator in the present context is that of one-particle energy (or
energy squared in the case of quantized relativistic bosonic
fields). The requirement of its self-adjointness renders the most
general set of boundary conditions, which may be further restricted
by some additional physical constraints.

   In the present paper we consider the Casimir effect with a quantized
charged massive scalar matter field in the background of an external
uniform magnetic field; both the quantized and external fields are
confined between two parallel plates, and the external field is
orthogonal to the plates. It should be noted that exactly this
problem has been studied more than a decade ago in
Refs.\cite{Cou,Eli,Ost}, however, results obtained there are
incomplete, and, therefore, somewhat misleading. In particular,
according to them, there is no room for the validation of the
aforementioned intuitive suggestion: the pressure in all
circumstances is negative, i.e. the plates are attracted. On the
contrary, we show that, even in the case of a weak magnetic field
and a small distance between the plates, the pressure is either
negative or positive, depending on the choice of a boundary
condition. A much more essential distinction from
Refs.\cite{Cou,Eli,Ost} is that, in the case of a strong magnetic
field and a large distance between the plates, the pressure is
positive, being independent of the choice of a boundary condition
and even of the distance between the plates.

 In the next section we consider in general the problem of the
self-adjointness for the operator of one-particle energy squared,
which is the same as that for the covariant Laplace operator. In Section
3 we discuss the vacuum energy density which is induced by an
external uniform magnetic field and compare the appropriate
expressions in the cases of the unbounded quantization volume and
the quantization volume bounded by two parallel plates. A choice of
boundary conditions for the quantized scalar field is considered in
Section 4. The expressions for the Casimir energy and force are
obtained in Section 5. The conclusions are drawn and discussed in
Section 6. We relegate some details of performing the infinite summation
over discrete eigenvalues to the Appendix.

\section{Self-adjointness of the Laplace operator}

Defining a scalar product as
$$
(\tilde{\chi},\chi)=\int\limits_{_D}d^3r\,\tilde{\chi}^{*}\chi\, ,
$$
we get, using integration by parts,
$$
(\tilde{\chi},\bnabla^2\chi)=(\bnabla^2\tilde{\chi},\chi)+
\int\limits_{\partial{D}}{\rm d}\boldsymbol{\sigma}[\tilde{\chi}^{*}(\bnabla\chi)-
(\bnabla\tilde{\chi})^{*}\chi], \eqno(1)
$$
where $\partial{D}$ is a two-dimensional surface bounding the
three-dimensional spatial region $D$, $\bnabla$ is the covariant
derivative involving both affine and bundle connections. The covariant Laplace
operator, $\bnabla^2$, is Hermitian (or symmetric in mathematical parlance),
$$
(\tilde{\chi},\bnabla^2\chi)=(\bnabla^2\tilde{\chi},\chi),\eqno(2)
$$
if
$$\int\limits_{\partial{D}}{\rm d}\boldsymbol{\sigma}\,[\tilde{\chi}^{*}(\bnabla\chi)-
(\bnabla\tilde{\chi})^{*}\chi] = 0. \eqno(3)$$ The latter condition
can be satisfied in various ways by imposing different boundary
conditions for $\chi$ and $\tilde{\chi}$. However, among the whole
variety, there may exist a possibility that a boundary condition for
$\tilde{\chi}$ is the same as that for $\chi$; then operator
$\bnabla^2$ is self-adjoint. The spectral theorem is valid for
self-adjoint operators only, and this allows one to construct appropriate unitary operator 
exponentials playing the key role in defining the dynamical evolution of quantum
systems, see, e.g., Ref.\cite{Akhi}.
  In the case of a disconnected noncompact boundary consisting of two components,
$\partial{D}^{(+)}$ and $\partial{D}^{(-)}$, condition (3) takes the
form,
$$\int\limits_{\partial{D}^{(+)}}{\rm d}\boldsymbol{\sigma}[\tilde{\chi}^{*}(\bnabla\chi)-
(\bnabla\tilde{\chi})^{*}\chi]-\int\limits_{\partial{D}^{(-)}}{\rm d}\boldsymbol{\sigma}[\tilde{\chi}^{*}(\bnabla\chi)-
(\bnabla\tilde{\chi})^{*}\chi] = 0, \eqno(4)$$ where normals to
surfaces $\partial{D}^{(+)}$ and $\partial{D}^{(-)}$ are chosen to
point in the same direction, i.e. outwards for $\partial D^{(+)}$ and inwards for $\partial D^{(-)}$. One can introduce coordinates
$\mathbf{r}=(x,y,z)$ in such a way, that $y$ and $z$ are tangential
to the boundary, while $x$ is normal to it, then the position of
$\partial{D}^{(\pm)}$ is identified with, say, $x=\pm{a}$. In this way we obtain
$$
(\tilde{\chi},\bnabla^2\chi)-(\bnabla^2\tilde{\chi},\chi)=
$$
$$
=\int{{\rm d}y}{{\rm d}z}\left(\tilde{\chi}^{*}|_{a}\nabla_{x}\chi|_{a}-
\nabla_{x}\tilde{\chi}^{*}|_{a}\chi|_{a}-\tilde{\chi}^{*}|_{-a}\nabla_{x}\chi|_{-a}+\nabla_{x}
\tilde{\chi}^{*}|_{-a}\chi|_{-a}\right)=$$
$$
=\frac{\rm i}{2a}\int{{\rm d}y}{{\rm d}z}(\tilde{\chi}^*_-|_{-a}\chi_-|_{-a}+\tilde{\chi}^*_+|_{a}\chi_+|_{a}-\tilde{\chi}^*_+
|_{-a}\chi_+|_{-a}-
\tilde{\chi}^*_-|_{a}\chi_-|_{a}), \eqno(5)
$$
where
$$
\chi_\pm =a\nabla_x\chi\pm {\rm i}\chi,\qquad \tilde{\chi}_\pm=a\nabla_x\tilde{\chi}\pm {\rm i}\tilde{\chi}.
$$
The integrand in (5) vanishes when the following condition is satisfied:
$$
\begin{pmatrix} \chi_-|_{-a} \\ \chi_+|_{a}\end{pmatrix}=
U\begin{pmatrix} \chi_+|_{-a} \\ \chi_-|_{a}\end{pmatrix},\quad
\begin{pmatrix} \tilde{\chi}_-|_{-a} \\ \tilde{\chi}_+|_{a}\end{pmatrix}=
U\begin{pmatrix} \tilde{\chi}_+|_{-a} \\ \tilde{\chi}_-|_{a}\end{pmatrix}, \eqno(6)
$$
where $U$ is a $U(2)$-matrix which is in general parametrized as
$$
U=e^{-{\rm i}\alpha}\begin{pmatrix}
u&v\\
-v^{*}&u^{*}
\end{pmatrix},\qquad
0<\alpha<\pi, \qquad |u|^{2}+|v|^{2}=1.\eqno(7)$$
Thus, the explicit form
of the boundary condition ensuring the self-adjointness of the
Laplace operator is
$$\{[1-e^{-{\rm i}\alpha}(u^{*}\pm{v})]a\nabla_{x}+{\rm i}[1+e^{-{\rm i}\alpha}(u^{*}\pm{v})]\}\chi|_{a}=$$
$$
=\{\mp[1-e^{-{\rm i}\alpha}(u\mp{v}^{*})]a\nabla_{x}\pm{\rm i}[1+e^{-{\rm i}\alpha}(u\mp{v}^{*})]\}\chi|_{-a}\eqno(8)
$$
(the same condition is for $\tilde{\chi}$).

 Similar considerations in a somewhat simplified form apply also to
the operator of the covariant momentum component in the normal to the boundary
direction,$-{\rm i}\nabla_{x}$:
$$
(\tilde{\chi},-{\rm i}\nabla_{x}\chi)=(-{\rm i}\nabla_{x}\tilde{\chi},\chi)-{\rm i}\int\limits_{\partial{D}}{\rm d}\sigma^{x}\,\tilde{\chi}^{*}\chi.\eqno
(9)$$ The explicit form of the boundary condition ensuring the
self-adjointness of $-{\rm i}\nabla_{x}$ is
$$
\chi|_{a}=\tilde{u}\chi|_{-a},\quad |\tilde{u}|^{2}=1 \eqno (10)
$$(the same condition is for $\tilde{\chi}$).

  In the one-dimensional case (when dimensions along the $y$ and $z$ axes are ignored), the deficiency
index is $\{2,2\}$ in the case of $\bnabla^{2}$, see Ref.\cite{Car}, and $\{1,1\}$ in the
case of $-{\rm i}\nabla_{x}$, see, e.g., Ref.\cite{Akhi}. The four real parameters from (7) and the
one real parameter from (10) are the self-adjoint extension
parameters. When the account is taken for dimensions along the
boundary, these parameters become arbitrary functions of $y$ and
$z$. However, in such a general case, conditions (8) and (10) cannot
be regarded as the ones determining the spectrum of the momentum in
the $x$-direction. Therefore, we assume that the self-adjoint
extension parameters are independent of $y$ and $z$.

Moreover, there are further restrictions which are due to physical reasons.
For instance, one may choose a one-parameter family of boundary
conditions, ensuring the self-adjointness of the momentum operator
in the $x$-direction, see (10), then the Laplace operator is surely
self-adjoint. However, this choice is too restrictive, with a lack
of physical motivation, and we shall follow another way. A solution
to the stationary Klein-Fock-Gordon equation, $\psi(\mathbf{r})$,
will be chosen in such a manner that its phase is independent of the
coordinate which is normal to the boundary. Then $U^{*}U=I$ which
results in
$$
v^{*}=-v,\eqno(11)
$$and the number of the self-adjoint extension parameters is
diminished to 3. The density of the conserved current in the
$x$-direction,
$$
j^{x}(\mathbf{r})=-{\rm i}[\psi^{*}(\nabla_{x}\psi)-(\nabla_{x}\psi)^{*}\psi],\eqno(12)
$$in this case vanishes at the boundary:
$$
j^{x}|_{a}=j^{x}|_{-a}=0,\eqno(13)
$$ and, thus, the matter is confined within the boundaries.

 A much more stringent restriction is condition $U^{2}=I$ which
results in
$$
(\psi,-\bnabla^2\psi)=(-{\rm i}\bnabla\psi,-{\rm i}\bnabla\psi),\eqno(14)
$$
meaning that the spectrum of the Laplace operator is non-positive-definite.
Condition (14) ensures that the values
of the one-particle energy squared exceed the value of the mass squared.

    \section{Induced vacuum energy density in \\ the magnetic field
    background}

  The operator of a charged massive scalar field which is quantized in a static background is presented in the
form
$$\hat{\Psi}(t,\mathbf{r})=\sum\!\!\!\!\!\!\!\int\limits_{\lambda}\frac{1}{\sqrt{2\omega_{\lambda}}}[e^{-\rm i\omega_{\lambda}t}\psi_{\lambda}(\mathbf{r})\hat{a}_{\lambda}
+e^{\rm i\omega_{\lambda}t}\psi^*_{\lambda}(\mathbf{r})\hat{b}^{\dag}_{\lambda}],\eqno(15)$$where
$\hat{a}^{\dag}_{\lambda}$ and $\hat{a}_{\lambda}$
($\hat{b}^{\dag}_{\lambda}$ and $\hat{b}_{\lambda}$) are the scalar
particle (antiparticle) creation and destruction operators,
satisfying commutation relations
$$
[\hat{a}_\lambda,\hat{a}_{\lambda'}^\dagger]_-=[\hat{b}_\lambda,\hat{b}_{\lambda'}^\dagger]_-=\left\langle \lambda|\lambda'\right\rangle;
$$
$\lambda$ is the set of parameters
(quantum numbers) specifying the state;
$\omega_{\lambda}>0$ is the energy of the state;
symbol $\sum\!\!\!\!\!\!\!\int\limits_{\lambda}\,$ denotes summation
over discrete and integration (with a certain measure) over
continuous values of $\lambda$; wave functions
$\psi_{\lambda}(\textbf{r})$ form a complete set of solutions to the
stationary Klein-Fock-Gordon equation
$$
[-\bnabla^{2}+m^2+\xi
R(\mathbf{r})]\psi_{\lambda}(\mathbf{r})=\omega_{\lambda}^{2}\psi_{\lambda}(\mathbf{r}),\eqno(16)
$$where $R(\mathbf{r})$ is the scalar curvature of space-time; $m$ is the particle mass. The temporal component of the energy-momentum tensor is given by expression
$$
\hat{T}^{00}=[{\partial_0}\hat{\Psi}^{\dag},{\partial_0}\hat{\Psi}]_{+}-\left[\frac{1}{4}({\partial_0}^2-\bnabla^2)+\xi(\bnabla^2+R^{00})\right][\hat{\Psi}^{\dag},\hat{\Psi}]_{+},\eqno(17)
$$where $R^{00}(\mathbf{r})$ is the temporal component of the Ricci
tensor. Thence, the formal expression for the vacuum energy
density is
$$
\varepsilon=<\texttt{vac}|\hat{T}^{00}|\texttt{vac}>=\sum\!\!\!\!\!\!\!\int\limits_{\lambda}\omega_{\lambda}\psi_{\lambda}^{*}(\textbf{r})\psi_{\lambda}(\textbf{r})+$$
$$+ \left[\left(\frac{1}{4}-\xi\right)\bnabla^2-\xi{R^{00}(\mathbf{r}})\right]\sum\!\!\!\!\!\!\!\int\limits_{\lambda}\omega_{\lambda}^{-1}\psi_{\lambda}^{*}(\textbf{r})\psi_{\lambda}(\textbf{r}).\eqno(18)
$$

It should be noted that, in general, the energy-momentum tensor and its vacuum expectation value remain dependent
on the coupling ($\xi$, see (16)) of the scalar field to the scalar
curvature of space-time even in the case of flat space-time, i.e.
when the Ricci tensor and the Ricci scalar vanish.
This is an evidence for some arbitrariness in the definition of the energy-momentum tensor for the scalar field in flat space-time. One can add term $\xi\nabla_{\rho}\theta^{\mu\nu\rho}$, where $\theta^{\mu\nu\rho}=-\theta^{\mu\rho\nu}$, to the canonically-defined energy-momentum tensor, $T^{\mu\nu}_{\rm can}$. The whole construction at $\xi=1/6$ is known as the improved energy-momentum tensor which is adequate for the implementation of conformal invariance in the $m=0$ case \cite{Che,Cal}. Physical observables are certainly independent of this arbitrariness, i.e. of $\xi$.

Let us consider the quantization of the charged massive scalar field in the background of a
uniform magnetic field $(\mathbf{B})$ in flat space-time
$(R=0,R^{00}=0)$, then the covariant derivative is defined as
$$\bnabla\hat{\Psi}=(\boldsymbol{\partial}-{\rm i}e\mathbf{A})\hat{\Psi},\quad
\bnabla\hat{\Psi}^{\dag}=(\boldsymbol{\partial}+{\rm i}e\mathbf{A})\hat{\Psi}^{\dag},\quad
\mathbf{B}=\boldsymbol{\partial}\times\mathbf{A},\eqno(19)$$ $e$ is
the particle charge. Directing the magnetic field along the
$x$-axis, $\mathbf{B}=(B,0,0)$, we choose the gauge with
$A^{x}=A^{y}=0,A^{z}=yB$. Then the solution to the Klein-Fock-Gordon
equation takes form
$$\psi_{knq}(\mathbf{r})=X_{k}(x)Y_{nq}(y)Z_{q}(z),\,
-\infty<k<\infty,\,-\infty<q<\infty,\,n=0,1,2,...\, ,\eqno(20)
$$ where
$$X_{k}(x)=(2\pi)^{-1/2}{\rm e}^{\rm{i}kx},\quad Z_{q}(z)=(2\pi)^{-1/2}{\rm e}^{\rm{i}qz}, \eqno(21)
$$ and
$$Y_{nq}(y)=\sqrt{\frac{|eB|^{1/2}}{2^{n}n!\pi^{1/2}}}\exp{\left[-\frac{|eB|}{2}\left(y+\frac{q}{eB}\right)^{2}\right]H_{n}\left[\sqrt{|eB|}\left(y+\frac{q}{eB}\right)\right]},\eqno(22)
$$ $H_n(w)$ is the Hermite polinomial. Wave functions $\psi_{knq}(\mathbf{r})$ satisfy the conditions of orthonormality
$$\int{{\rm d}^{3}r}\,\psi^{*}_{knq}(\mathbf{r})\psi_{k'n'q'}(\mathbf{r})=\delta(k-k')\delta_{nn'}\delta(q-q'),\eqno(23)
$$and completeness
$$\int\limits_{-\infty}^{\infty}{\rm{d}}k\int\limits_{-\infty}^{\infty}{\rm{d}}q\sum\limits_{n=0}^{\infty}\psi^{*}_{knq}(\mathbf{r})\psi_{knq}(\mathbf{r'})=\delta^3(\mathbf{r}-\mathbf{r'}).\eqno(24)
$$
The one-particle energy spectrum (Landau levels) is the following
$$\omega_{kn}=\sqrt{|eB|(2n+1)+k^{2}+m^{2}}.\eqno(25)
$$With the use of relations
$$\int\limits_{-\infty}^{\infty}{\rm{d}}q\,{Y^{2}_{nq}(y)}=|eB|,\quad
\int\limits_{-\infty}^{\infty}{\rm{d}}q\,{\partial_{y}^{2}Y^{2}_{nq}(y)}=0,\eqno(26)
$$the formal expression for the vacuum energy density in the present
case is readily obtained:
$$\varepsilon^{\infty}=\frac{|eB|}{(2\pi)^{2}}\int\limits_{-\infty}^{\infty}{\rm d}k\sum\limits_{n=0}^{\infty}\omega_{kn},\eqno(27)
$$
where the superscript indicates that the external magnetic field fills the whole (infinite) space; note
that dependence on $\xi$ has disappeared (just owing to the second
relation in (26)). The integral and the sum in (27) are divergent
and require regularization and renormalization. This problem has
been solved long ago by Weisskopf \cite{Wei} (see also Ref.\cite{Schw}), and we just list here his
result
$$\varepsilon^{\infty}_{\rm ren}=-\frac{1}{(4\pi)^{2}}\int\limits_{0}^{\infty}\frac{{\rm d}\tau}{\tau}{\rm e}^{-\tau}\left[\frac{eBm^{2}}{\tau\sinh\left(\frac{eB\tau}{m^{2}}\right)}-\frac{m^{4}}{\tau^2}+\frac{1}{6}e^{2}B^{2}\right];\eqno(28)
$$note that the renormalization procedure includes subtraction at
$\mathbf{B}=0$ and renormalization of the charge.

  Let us turn now to the quantization of the charged massive scalar field
in the background of a uniform magnetic field in spatial region
$D$ bounded by two parallel surfaces $\partial{D}^{(+)}$ and
$\partial{D}^{(-)}$; the position of $\partial{D}^{(\pm)}$ is
identified with $x=\pm{a}$, and the magnetic field is orthogonal to
the boundary. Then the solution to the Klein-Fock-Gordon equation
takes form
$$\psi_{lnq}(\mathbf{r})=X_{l}(x)Y_{nq}(y)Z_{q}(z),
l=0,\pm1,\pm2,...,-\infty<q<\infty,n=0,1,2,...,\eqno(29)
$$where $Y_{nq}(y)$ and $Z_{q}(z)$ are the same as in the previous case,
while $X_{l}(x)$ is the real solution to equation
$$(\partial_{x}^{2}+k_{l}^{2})X_{l}(x)=0,\eqno (30)
$$and the discrete spectrum of $k_{l}$ is determined by the
boundary condition for $X_{l}(x)$, see (8) with (11):
$$\{[1-{\rm e}^{-{\rm i}\alpha}(u^{*}\pm{v})]a\partial_{x}+{\rm i}[1+{\rm e}^{-{\rm i}\alpha}(u^{*}\pm{v})]\}X_{l}|_{a}=$$
$$=\{\mp[1-{\rm e}^{{\rm{-i}}\alpha}(u\pm{v})]a\partial_{x}\pm{\rm i}[1+{\rm e}^{-{\rm i}\alpha}(u\pm{v})]\}X_{l}|_{-a}\quad (v^*=-v).\eqno(31)
$$ Wave functions $\psi_{lnq}(\mathbf{r})$ satisfy the conditions of orthonormality
$$\int\limits_{D}{{\rm d}^{3}r}\,\psi^{*}_{lnq}(\mathbf{r})\psi_{l'n'q'}(\mathbf{r})=\delta_{ll'}\delta_{nn'}\delta(q-q'),\eqno(32)$$
and completeness
$$\int\limits_{-\infty}^{\infty}{\rm d}q\sum\limits_{l=-\infty}^{\infty}\sum\limits_{n=0}^{\infty}\psi^{*}_{lnq}(\mathbf{r})\psi_{lnq}(\mathbf{r'})=\delta^3(\mathbf{r}-\mathbf{r'}).\eqno(33)$$
The formal expression for the vacuum energy density appears to be
$\xi$-dependent, cf. (27),
$$\varepsilon=\frac{|eB|}{2\pi}\sum\limits_{l=-\infty}^{\infty}\sum\limits_{n=0}^{\infty}\,\left[\omega_{ln}+\left(\frac{1}{4}-\xi\right)\omega_{ln}^{-1}\partial_{x}^{2}\right]X_{l}^{2}(x),\eqno(34)
$$where
$$\omega_{ln}=\sqrt{|eB|(2n+1)+k_{l}^{2}+m^{2}}.\eqno(35)$$

\section{Choice of boundary conditions for \\ the Casimir effect}

 The Casimir energy is defined as the induced vacuum energy per unit area of the  boundary surface:
$$\frac{E}{S}=\int\limits_{-a}^{a}{\rm{d}}x\,\varepsilon.\eqno(36)
$$In view of the normalization condition, we obtain the formal expression for the
Casimir energy
$$\frac{E}{S}=\frac{|eB|}{2\pi}\sum\limits_{l=-\infty}^{\infty}\sum\limits_{n=0}^{\infty}\left[\omega_{ln}+\left(\frac{1}{4}-\xi\right)\omega_{ln}^{-1}I_{l}\right],\eqno(37)
$$where
$$I_{l}=\int\limits_{-a}^{a}{\rm{d}}x\,\partial_{x}^{2}X_{l}^{2}(x).\eqno(38)
$$ By rewriting the integrand in (38) as $2X_{l}(\partial^{2}_{x}X_{l})+2(\partial_{x}X_{l})^{2}$, one immediately recognizes
that condition (14) ensures the vanishing of $I_{l}$. Thus, this
condition guarantees that the Casimir energy in flat space is
independent of the $\xi$-parameter, as it should be expected for a
physically meaningful quantity.

  Without loss of generality, one may take $X_{l}(x)$ in the form
$$X_{l}(x)=\frac{1}{\sqrt{a}}\sin(k_{l}x+\delta_{l}), \eqno(39)
$$with phase $\delta_{l}$, as well as momentum $k_{l}$, being
determined from boundary condition (31). Substituting (39) into
(38), one gets
$$I_{l}=\frac{kl}{a}\cos(2\delta_{l})\sin(2k_{l}a).\eqno(40)$$
 The condition of the vanishing of $I_{l}$ (40) restricts the number and the range of self-adjoint
extension parameters.

  Let us consider the following cases:

  I. $v=0$, then
$$U=\exp(-{\rm i}\alpha{I}+{\rm i}\tilde{\alpha}\sigma_{3}),\quad
0<\alpha<\pi,\quad 0<\tilde{\alpha}<2\pi, \eqno(41)
$$
where parametrization $\texttt{Re}{u}=\cos{\tilde{\alpha}},
\texttt{Im}{u}=\sin{\tilde{\alpha}}$ is used;

  II. $\texttt{Re}u=0$ and $\alpha=\frac{\pi}{2}$, then
$$U=\sigma_{1}\cos{\beta}+\sigma_{3}\sin{\beta},\quad 0<\beta<2\pi,\eqno(42)
$$where parametrization $\texttt{Im}u=\sin{\beta}, \texttt{Im}v=\cos{\beta}$ is used;

  III. $\texttt{Im}u=0$, then
$$U=\exp(-{\rm i}\alpha{I}+{\rm i}\tilde{\beta}\sigma_{1}), \quad 0<\alpha<\pi,\quad -\pi<\tilde{\beta}<\pi,\eqno(43)
$$where parametrization $\texttt{Re}u=\cos{\tilde{\beta}},
\texttt{Im}v=\sin{\tilde{\beta}}$ is used; here $\sigma_{j}$ are the
Pauli matrices, $I$ is the unity matrix, the endpoints of the ranges
are identified.

Boundary condition (31) takes the form
$$k_{l}a\cos(\pm{k_{l}a}+\delta_{l})\sin\left[\frac{1}{2}(\alpha\pm\tilde{\alpha})\right]\pm\sin(\pm{k_{l}a}+\delta_{l})\cos\left[\frac{1}{2}(\alpha\pm\tilde{\alpha})\right]=0\eqno(44)
$$
in case I,
$$
\left\{\begin{array}{l} \cos(k_{l}a+\delta_{l})\cos\left(\frac{1}{2}\beta - \frac{\pi}{4} \right)+\cos(-k_{l}a+\delta_{l})\sin\left(\frac{1}{2}\beta - \frac{\pi}{4}\right)=0 \\ [3 mm]
\sin(k_{l}a+\delta_{l})\sin\left(\frac{1}{2}\beta - \frac{\pi}{4}\right) + \sin(-k_{l}a+\delta_{l})\cos\left(\frac{1}{2}\beta - \frac{\pi}{4}\right)=0\end{array}
\right\}\eqno(45)
$$
in case II,
$$
\left\{\begin{array}{l} k_{l}a\cos(k_{l}a)\sin\left[\frac{1}{2}(\alpha-\tilde{\beta})\right]+\sin(k_{l}a)\cos\left[\frac{1}{2}(\alpha-\tilde{\beta})\right]=0 \\ [3 mm]
k_{l}a\sin(k_{l}a)\sin\left[\frac{1}{2}(\alpha+\tilde{\beta})\right]-\cos(k_{l}a)\cos\left[\frac{1}{2}(\alpha+\tilde{\beta})\right]=0\end{array}
\right\}\eqno(46)
$$
in case III.

   In case I, condition $\alpha=\tilde{\alpha}=0$ corresponds to the Dirichlet boundary condition $\left(X_{l}|_{-a}=
X_{l}|_{a}=0\right)$ which yields the spectrum
$$\delta_{l}=0, \quad  k_{l}=l\frac{\pi}{a}, \quad l=1,2,...,\eqno(47)$$
$$\delta_{l}=\frac{\pi}{2},\quad k_{l}=\left(l+\frac{1}{2}\right)\frac{\pi}{a},\quad l=0,1,2,... \,.\eqno(48)$$
  In case I, condition $\alpha=\tilde{\alpha}-\pi=0$ corresponds to the Neumann boundary condition $\left(\partial_{x}X_{l}|_{-a}=
\partial_{x}X_{l}|_{a}=0\right)$ which yields the spectrum
$$\delta_{l}=(l-1)\frac{\pi}{2},\quad k_{l}=l\frac{\pi}{2a},\quad l=1,2,... \,.\eqno(49)$$
  In case I, condition $\alpha=\tilde{\alpha}=\frac{\pi}{2}$ corresponds to the mixed Dirichlet-Neumann boundary
condition $\left(X_{l}|_{-a}=\partial_{x}X_{l}|_{a}=0\right)$ which
yields the spectrum
$$\delta_{l}=\left(l+\frac{1}{2}\right)\frac{\pi}{2},\quad k_{l}=\left(l+\frac{1}{2}\right)\frac{\pi}{2a},\quad l=0,1,2,... \,.\eqno(50)$$
The mixed Neumann-Dirichlet boundary condition (
$\partial_{x}X_{l}|_{-a}=X_{l}|_{a}=0$; \\ $\alpha=\tilde{\alpha}-\pi=\frac{\pi}{2}$)
yields the same spectrum as in (50) with the opposite phase.

  In case II the spectrum is determined for the whole range of $\beta$:
$$\delta_{l}=\frac{\pi}{4},\quad k_{l}=l\frac{\pi}{a}+\frac{\beta}{2a},\quad l=0,\pm{1},\pm{2},... \,.\eqno(51)$$

   The boundary condition in case III, see (46), is independent of phase
$\delta_{l}$; condition $\alpha=\tilde{\beta}=\frac{\pi}{2}$
corresponds to the periodicity boundary condition
($X_{l}|_{-a}=X_{l}|_{a}$) yielding the spectrum
$$k_{l}=l\frac{\pi}{a},\quad l=0,\pm{1},\pm{2},...\, ,\eqno(52)
$$while condition $\alpha=-\tilde{\beta}=\frac{\pi}{2}$ corresponds to the
antiperiodicity boundary condition ($X_{l}|_{-a}=-X_{l}|_{a}$)
yielding the spectrum
$$k_{l}=\left(l+\frac{1}{2}\right)\frac{\pi}{a},\quad l=0,\pm{1},\pm{2},... \,.\eqno(53)
$$Results (52) and (53) are also obtained in case II at $\beta=0$ and
$\beta=\pi$ , respectively, without any restriction on phase
$\delta_{l}$.

  The boundary conditions yielding the spectra of $k_{l}$ in (47)-(53) ensure the vanishing of
$I_{l}$ (40). This is a complete list of conditions giving $U^{2}=I$.

It should be noted that the self-adjointness of the momentum in the $x$-direction, see (9) and (10), implies
the vanishing of quantity
$$
\tilde{I}_{l}=\int\limits_{-a}^{a}{\rm{d}}x\partial_{x}X_{l}^{2}(x)=\frac{1}{a}\sin(2\delta_{l})\sin(2k_{l}a).\eqno(54)
$$ Thus, the operator $-{\rm i}\partial_{x}$ is not self-adjoint under the mixed (Dirichlet-Neumann or Neumann-Dirichlet)
boundary condition, when (50) holds, and under the one-parameter
family of boundary conditions, when (51) holds, unless
$\beta=0,\pi$.

     \section{Casimir energy and force}

 Employing the boundary conditions specified in the previous section, we obtain the following expression for the Casimir energy
$$
\frac{E}{S}=\frac{|eB|}{2\pi}\sum\limits_{l=-\infty}^{\infty}\sum\limits_{n=0}^{\infty}\omega_{ln}.\eqno(55)
$$
The expression is ill-defined, as was already noted, since infinite
sums in (55) are divergent. To tame the divergence, a factor
containing the regularization parameter is inserted in (55).

  Let us perform calculations for case II of the previous section, when the boundary condition is given by (45)
and the spectrum of $k_{l}$ is given by (51). The summation over $l$
is made with the use of the following version of the Abel-Plana formula
which is derived in Appendix (see also Ref.\cite{Bel}):
\newpage
$$\sum\limits_{l=-\infty}^{\infty}f\left[\left(\pi{l}+\frac{\beta}{2}\right)^{2}\right]=\frac{1}{\pi}\int\limits_{-\infty}^{\infty}{\rm{d}}\mu{f(\mu^{2})}-$$
$$-\frac{\rm i}{\pi}\int\limits_{0}^{\infty}{\rm d}\nu\{f[(-{\rm i}\nu)^{2}]-f[({\rm i}\nu)^{2}]\}\frac{\cos{\beta}-{\rm e}^{-2\nu}}{\cosh{2\nu}-\cos{\beta}},\eqno (56)
$$where $f(w^{2})$ as a function of complex variable $w$ is decreasing
sufficiently fast at large distances from the origin of the complex
$w$-plane. The regularization in the last integral on the right-hand
side of (56) can be removed; then
$${\rm i}\{f[(-{\rm i}\nu)^{2}]-f[({\rm i}\nu)^{2}]\}=\frac{|eB|}{\pi}\sum\limits_{n=0}^{\infty}\sqrt{\left(\frac{\nu}{a}\right)^{2}-|eB|(2n+1)-m^{2}} \eqno(57)
$$
with the range of $\nu$ restricted to $\nu>a\sqrt{|eB|(2n+1)+m^{2}}$.
Introducing variable $k=\mu/a$ in the first integral on the right-hand side of (56), one immediately recognizes
that this integral is the same as quantity $\varepsilon^{\infty}$
(27) multiplied by $2a$. Hence, the problem of regularization and
removal of the divergency in expression (55) is reduced to that in
the case of no boundaries, when the magnetic field fills the whole
space. This is owing to the Abel-Plana formula (56) which
effectively reduces the contribution of the boundaries to the term
(last integral on the right-hand side) that is free of the divergency.
Thus we obtain the following expression for the renormalized Casimir
energy:
$$
\frac{E_{\rm ren}}{S}=2a\varepsilon^{\infty}_{\rm ren}-
\frac{|eB|}{\pi^{2}a}\sum\limits_{n=0}^{\infty}\int\limits_{aM_{n}}^{\infty}{\rm d}\nu\sqrt{\nu^{2}-a^{2}M_{n}^{2}}\frac{\cos{\beta}-{\rm e}^{-2\nu}}{\cosh{2\nu}-\cos{\beta}},\eqno(58)
$$ where $\varepsilon^{\infty}_{\rm ren}$ is given by (28) and
$$M_{n}=\sqrt{|eB|(2n+1)+m^{2}}. \eqno (59)$$

  In particular, in the case of the periodicity boundary condition when the spectrum is given by (52),
we get $\beta=0$ and
$$
\frac{E_{\rm ren}}{S}=2a\varepsilon^{\infty}_{\rm ren}-
\frac{2|eB|}{\pi^{2}a}\sum\limits_{n=0}^{\infty}\int\limits_{aM_{n}}^{\infty}{\rm d}\nu\frac{\sqrt{\nu^{2}-a^{2}M_{n}^{2}}}{{\rm e}^{2\nu}-1},\eqno(60)
$$while, in the case of the antiperiodicity boundary condition when
the spectrum is given by (53), we get $\beta=\pi$ and
$$
\frac{E_{\rm ren}}{S}=2a\varepsilon^{\infty}_{\rm ren}+
\frac{2|eB|}{\pi^{2}a}\sum\limits_{n=0}^{\infty}\int\limits_{aM_{n}}^{\infty}{\rm d}\nu\frac{\sqrt{\nu^{2}-a^{2}M_{n}^{2}}}{{\rm e}^{2\nu}+1}.\eqno(61)
$$

  The case of the mixed (either Dirichlet-Neumann or
Neumann-Dirichlet) boundary condition when the spectrum is given by
(50) is obtained from (61) by an appropriate rescaling of the
integration variable
$$
\frac{E_{\rm ren}}{S}=2a\varepsilon^{\infty}_{\rm ren}+
\frac{2|eB|}{\pi^{2}a}\sum\limits_{n=0}^{\infty}\int\limits_{aM_{n}}^{\infty}{\rm d}\nu\frac{\sqrt{\nu^{2}-a^{2}M_{n}^{2}}}{{\rm e}^{4\nu}+1}.\eqno(62)
$$

  The case of the Dirichlet boundary condition which is given by (44) at $\alpha=\tilde{\alpha}=0$ deserves a special attention. The modes
in this case are divided into two series of opposite parity. For the
modes of even parity, see (48), using
$$
\sum\limits_{l=0}^{\infty}f\left[\left(\pi{l}+\frac{\pi}{2}\right)^{2}\right]=\frac{1}{2}\sum\limits_{l=-\infty}^{\infty}f\left[\left(\pi{l}+\frac{\pi}{2}\right)^{2}\right],\eqno
(63)$$ we obtain
$$
\frac{E}{S}|_{\rm even}=a\varepsilon^{\infty}+
\frac{|eB|}{\pi^{2}a}\sum\limits_{n=0}^{\infty}\int\limits_{aM_{n}}^{\infty}{\rm d}\nu\frac{\sqrt{\nu^{2}-a^{2}M_{n}^{2}}}{{\rm e}^{2\nu}+1}.
$$For the modes of odd parity, see (47), using
$$
\sum\limits_{l=1}^{\infty}f\left[\left(\pi l\right)^{2}\right]=\frac{1}{2}\sum\limits_{l=-\infty}^{\infty}f\left[\left(\pi l\right)^{2}\right]
-\frac{1}{2}f(0),\eqno (64)$$ we obtain
$$
{\frac{E}{S}}|_{\rm odd}=a\varepsilon^{\infty}-
\frac{|eB|}{\pi^{2}a}\sum\limits_{n=0}^{\infty}\int\limits_{aM_{n}}^{\infty}{\rm d}\nu\frac{\sqrt{\nu^{2}-a^{2}M_{n}^{2}}}{{\rm e}^{2\nu}-1}-\frac{|eB|}{4\pi}\sum\limits_{n=0}^{\infty}M_{n}.
$$ Summing the contributions of both parities, we get
$$
\frac{E}{S}=2a\varepsilon^{\infty}-
\frac{2|eB|}{\pi^{2}a}\sum\limits_{n=0}^{\infty}\int\limits_{aM_{n}}^{\infty}{\rm d}\nu\frac{\sqrt{\nu^{2}-a^{2}M_{n}^{2}}}{{\rm e}^{4\nu}-1}-\frac{|eB|}{4\pi}\sum\limits_{n=0}^{\infty}M_{n}.
$$Hence, by removing the divergency in the same manner as in the case of no boundaries, we arrive at the expression
containing infinities,
$$
\frac{\tilde{E}_{\rm ren}}{S}=2a\varepsilon^{\infty}_{\rm ren}-
\frac{2|eB|}{\pi^{2}a}\sum\limits_{n=0}^{\infty}\int\limits_{aM_{n}}^{\infty}{\rm d}\nu\frac{\sqrt{\nu^{2}-a^{2}M_{n}^{2}}}{{\rm e}^{4\nu}-1}-\frac{|eB|}{4\pi}\sum\limits_{n=0}^{\infty}M_{n};\eqno(65)
$$the last sum in (65) is divergent.

  The same situation is encountered in the case of the Neumann boundary condition, see (44) at $\alpha=\tilde{\alpha}-\pi=0$, when the
spectrum is given by (49). Using (64), we arrive finally at the same
expression as (65). Moreover, if a zero mode of even parity,
$X_{0}=\frac{1}{\sqrt{2a}}$, is added, then the divergent sum enters
with the opposite sign:
$$
\frac{\tilde{E}_{\rm ren}}{S}=2a\varepsilon^{\infty}_{\rm ren}-
\frac{2|eB|}{\pi^{2}a}\sum\limits_{n=0}^{\infty}\int\limits_{aM_{n}}^{\infty}{\rm d}\nu\frac{\sqrt{\nu^{2}-a^{2}M_{n}^{2}}}{{\rm e}^{4\nu}-1}+\frac{|eB|}{4\pi}\sum\limits_{n=0}^{\infty}M_{n}.\eqno(66)
$$

  However, a physically measurable characteristics of the Casimir effect is the Casimir force which is
defined as the force per unit area of the boundary, or pressure:
$$
F=-\frac{1}{2}\frac{\partial}{\partial a}\frac{E_{\rm ren}}{S}.\eqno (67)
$$The divergent pieces of the Casimir energy in (65) and (66) do not
contribute to the force, since they are independent of $a$. Hence,
we obtain the following expression for the Casimir force in the case
of either Dirichlet or Neumann boundary condition:
$$
F=-\varepsilon^{\infty}_{\rm ren}-
\frac{|eB|}{\pi^{2}a^{2}}\sum\limits_{n=0}^{\infty}\int\limits_{aM_{n}}^{\infty}{\rm d}\nu\frac{\nu^{2}}{\sqrt{\nu^{2}-a^{2}M_{n}^{2}}}\frac{1}{{\rm e}^{4\nu}-1}.\eqno(68)
$$In the case of the mixed boundary condition the Casimir force is
$$
F=-\varepsilon^{\infty}_{\rm ren}+
\frac{|eB|}{\pi^{2}a^{2}}\sum\limits_{n=0}^{\infty}\int\limits_{aM_{n}}^{\infty}{\rm d}\nu\frac{\nu^{2}}{\sqrt{\nu^{2}-a^{2}M_{n}^{2}}}\frac{1}{{\rm e}^{4\nu}+1},\eqno(69)
$$while in the cases of the periodicity and antiperiodicity boundary
conditions it takes forms
$$
F=-\varepsilon^{\infty}_{\rm ren}-
\frac{|eB|}{\pi^{2}a^{2}}\sum\limits_{n=0}^{\infty}\int\limits_{aM_{n}}^{\infty}{\rm d}\nu\frac{\nu^{2}}{\sqrt{\nu^{2}-a^{2}M_{n}^{2}}}\frac{1}{{\rm e}^{2\nu}-1}\eqno(70)
$$ and
$$
F=-\varepsilon^{\infty}_{\rm ren}+
\frac{|eB|}{\pi^{2}a^{2}}\sum\limits_{n=0}^{\infty}\int\limits_{aM_{n}}^{\infty}{\rm d}\nu\frac{\nu^{2}}{\sqrt{\nu^{2}-a^{2}M_{n}^{2}}}\frac{1}{{\rm e}^{2\nu}+1},\eqno(71)
$$respectively.

It should be noted that the integrals in (70) and (71) can be taken after expanding the last factors
as $\sum\limits_{j=1}^{\infty}(\pm1)^{j-1}{\rm e}^{-2j\nu}$.
In this way, we obtain the following expressions for the Casimir
energy
$$
\frac{E_{\rm ren}}{S}=2a\varepsilon^{\infty}_{\rm ren}\mp\frac{|eB|}{\pi^{2}}\sum\limits_{n=0}^{\infty}M_{n}\sum\limits_{j=1}^{\infty}(\pm1)^{j-1}\frac{1}{j}K_{1}(2jaM_{n})\eqno(72)
$$and the Casimir force
$$
F=-\varepsilon^{\infty}_{\rm ren}\mp\frac{|eB|}{\pi^{2}}\sum\limits_{n=0}^{\infty}M_{n}^{2}\sum\limits_{j=1}^{\infty}(\pm1)^{j-1}\left[K_{0}(2jaM_{n})+\frac{1}{2jaM_{n}}K_{1}(2jaM_{n})\right],\eqno(73)
$$where the upper (lower) sign corresponds to the periodicity
(antiperiodicity) boundary condition, $K_{\rho}(t)$ is the Macdonald
function of order $\rho$. Similarly, we obtain an alternative to
(68) and (69) representation for the Casimir force in the cases of
the Dirichlet or Neumann boundary condition (upper sign) and the
mixed boundary condition (lower sign):
$$
F=-\varepsilon^{\infty}_{\rm ren}\mp\frac{|eB|}{\pi^{2}}\sum\limits_{n=0}^{\infty}M_{n}^{2}\sum\limits_{j=1}^{\infty}(\pm1)^{j-1}\left[K_{0}(4jaM_{n})+\frac{1}{4jaM_{n}}K_{1}(4jaM_{n})\right].\eqno(74)
$$

  Finally, we present the Casimir force in the case of the one-parameter family of boundary conditions given by (45):
$$
F=-\varepsilon_{\rm ren}^\infty-\frac{|eB|}{2\pi^2a^2}\sum\limits_{n=0}^{\infty}\int\limits_{aM_n}^{\infty}{\rm d}\nu
\frac{\nu^2}{\sqrt{\nu^2-a^2M_n^2}}\frac{\cos \beta-{\rm e}^{-2\nu}}{{\rm cosh}2\nu-\cos\beta}.\eqno(75)
$$

\section{Conclusion and discussion}

In the present paper, we have considered the influence of a
background uniform magnetic field and boundary conditions on the
vacuum of a quantized charged massive scalar matter field confined
between two parallel plates separated by distance 2a. The magnetic
field is assumed to be directed orthogonally to the plates, then
the covariant Laplace operator is self-adjoint under a set of
boundary conditions depending on four arbitrary functions of two
coordinates which are tangential to the plates. Ignoring this
functional dependence and imposing a condition that the flow of
quantized matter outside the bounding plates is absent, see (13),
we arrive at the set of boundary conditions depending on three
arbitrary parameters, see (7) with (11). A further restriction,
see (14), is due to a physical requirement that the operator of
one-particle energy squared be positive definite, which makes the
Casimir effect to be independent of $\xi$ -- the coupling of the
scalar field to the scalar curvature of space-time. This reduces
the $U(2)$-matrix defining the boundary condition, see (6), to the
form given by (42), rendering finally the set of boundary
conditions depending on one parameter, $\beta$, in the range
$0<\beta<2\pi$ with the endpoints identified; under these
circumstances the Casimir force is shown to take the form of (75).
In particular, the Casimir force in the cases of the periodicity
and the antiperiodicity boundary conditions is obtained at
$\beta=0$ and $\beta=\pi$, respectively, see (70) and (71), or
alternatively (73); while the Casimir force in the cases of the
Dirichlet (or Neumann) and the mixed Dirichlet-Neumann (or
Neumann-Dirichlet) boundary conditions is obtained from the two
preceding ones by change $a\rightarrow 2a$, see (68) and (69), or
alternatively (74).

In the limit of a weak magnetic field, $|B| \ll m^2|e|^{-1}$, one has (see Ref.\cite{Wei})
$$
\varepsilon_{\rm ren}^\infty=-\frac{7}{8}\,\frac{1}{720\pi^2}\,\frac{e^4B^4}{m^4}.\eqno(76)
$$
Thus, at $|B|\rightarrow 0$ the first term on the right-hand side of (75) vanishes, and, substituting the sum in the remaining part there by integral $\int\limits_{0}^{\infty}{\rm d}n$, we get
$$
F|_{B=0}=-\frac{1}{2\pi^2a^4}\int\limits_{am}^{\infty}{\rm d}\nu
\nu^2{\sqrt{\nu^2-a^2m^2}}\,\frac{\cos \beta-{\rm e}^{-2\nu}}{{\rm cosh}2\nu-\cos\beta},\eqno(77)
$$
which in the limits of large and small distances between the plates takes the forms:
\newpage
$$
F|_{B=0}=-\frac{{m}^{5/2}}{4(\pi a)^{3/2}}\left\{\cos\beta {\rm e}^{-2am}\left[1+{\rm O}\left(\frac{1}{am}\right)\right]+\right.
$$
$$\left.+\frac{\cos\,2\beta}{2\sqrt{2}}{\rm e}^{-4am}\left[1+{\rm O}\left(\frac{1}{am}\right)\right]+{\rm O}({\rm e}^{-6am})\right\},\quad am\gg 1\eqno(78)
$$
and
$$
F|_{B=0}=-\frac{1}{2\pi^2a^4}\int\limits_{0}^{\infty}{\rm d}\nu\,\nu^3\frac{\cos\beta-{\rm e}^{-2\nu}}{{\rm cosh}2\nu-\cos\beta}=
-\frac{\pi^2}{8a^4}\left[\frac{1}{30}-\frac{\beta^2}{(2\pi)^2}\left(1-\frac{\beta}{2\pi}\right)^2\right],
$$
$$
am\ll 1.\eqno(79)
$$
The standard results for the Casimir force in the case of the massless charged scalar field are obtained from (79) at $\beta=0$ (periodicity boundary condition), see, e.g., Ref.\cite{Bor},
$$
F|_{B=0\atop m=0}=-\frac{\pi^2}{240}\,\frac{1}{a^4}\eqno(80)
$$
and at $\beta=\pi$ (antiperiodicity boundary condition),
$$
F|_{B=0\atop m=0}=\frac{7}{8}\,\frac{\pi^2}{240}\,\frac{1}{a^4};\eqno(81)
$$
the results for $\left.F\right|_{B=0\atop m=0}$ under the Dirichlet (or Neumann)
and the mixed Dirichlet-Neumann (or Neumann-Dirihlet) conditions are obtained from (80) and (81), respectively, by changing $a\rightarrow 2a$.

In the limit of a strong magnetic field, $|B|\gg m^2|e|^{-1}$, one has (see, e.g., Ref.\cite{Du})
$$
\varepsilon_{\rm ren}^\infty=-\frac{e^2B^2}{96\pi^2}\ln\frac{2 |{\rm e}B|}{m^2},\eqno(82)
$$
while the remaining piece of the force is
$$
\left(F+\varepsilon_{\rm ren}^\infty\right)|_{m=0}=-\frac{|eB|}{2\pi^2a^2}\sum\limits_{n=0}^{\infty}
\int\limits_{a\sqrt{|eB|(2n+1)}}^{\infty}{\rm d}\nu\frac{\nu^2}{\sqrt{\nu^2-a^2|eB|(2n+1)}}\times
$$
$$
\times \frac{\cos\beta-{\rm e}^{-2\nu}}
{{\rm cosh}2\nu-\cos\beta}.\eqno(83)
$$
The latter expression in the limits of large and small distances between the plates takes the forms:
$$
\left(F+\varepsilon_{\rm ren}^{\infty}\right)|_{m=0}=-\frac{|eB|^{7/4}}{2\pi^{3/2}a^{1/2}}\sum\limits_{n=0}^{\infty}
(2n+1)^{3/4}\biggl\{\cos\beta {\rm e}^{-2a\sqrt{|eB|(2n+1)}}\biggl[1+
$$
$$
+{\rm O}\biggl(\frac{1}{a\sqrt{|eB|(2n+1)}}\biggr)\biggr]+\frac{\cos 2\beta}{\sqrt{2}}{\rm e}^{-4a\sqrt{|eB|(2n+1)}}
\biggl[1+
$$
$$
+{\rm O}\biggl(\frac{1}{a\sqrt{|eB|(2n+1)}}\biggr)\biggr]+{\rm O}\left({\rm e}^{-6a\sqrt{|eB|(2n+1)}}\right)\biggr\},
\,\,a\sqrt{|eB|}\gg 1\eqno(84)
$$
and
$$
(F+\varepsilon_{\rm ren}^\infty)|_{m=0}=-\frac{\pi^2}{8a^4}\left[\frac{1}{30}-\frac{\beta^2}{(2\pi)^2}\left(1-\frac{\beta}{2\pi}\right)^2\right],\,\,a\sqrt{|eB|}\ll 1.\eqno(85)
$$

We can conclude that the Weisskopf term, $\varepsilon_{\rm
ren}^\infty$ (28), is dominating at a relatively large separation
of the plates, $2a\gg 2m^{-1}$, at a nonweak magnetic field. In
this case the Casimir force, $F\approx -\varepsilon_{\rm
ren}^\infty$, is repulsive (the pressure from the vacuum is
positive), being independent of the choice of boundary conditions
at the plates, as well as of the distance between the plates. In
the opposite case of a relatively small separation of the plates,
$2a\ll 2m^{-1}$, at a sufficiently weak magnetic field, $|B| \ll
m^2|e|^{-1}$, the Weisskopf term is negligible, and the Casimir
force, being power dependent on the distance between the plates as
$(2a)^{-4}$, see (79), is either attractive of repulsive,
depending on the choice of boundary conditions. A numerical
analysis of a regime which is intermediate between the two above
will be considered elsewhere.

Let us compare our results with those of our predecessors, see Refs.\cite{Cou,Eli,Ost}. It should be noted in the first place that these authors have disregarded the dependence on the choice of boundary conditions, restricting themselves to the choice of the Dirichlet one. Secondly, they present expressions for the Casimir energy only, and the latter lacks immediate physical meaning. In particular, the author of Ref.\cite{Ost} uses the Abel-Plana formula to obtain the Casimir energy, but simply drops without any explanation (see transition from (18) to (19) in Ref.\cite{Ost}) a divergency given by the last sum in (65); fortunately, this divergency has no effect on the Casimir force, as we have discussed in Section 5. An approach of the authors of Refs.\cite{Cou,Eli} is different, and by using some analytical methods of regularization they obtain a finite piece of the Casimir energy, which is actually given by before the last sum in (65). A common fault of Refs.\cite{Cou, Eli, Ost} is that the Weisskopf-term contribution to the Casimir energy, $2a\varepsilon_{\rm ren}^\infty$, is completely ignored.

Finally, let us discuss the contribution of the quantized
electromagnetic matter field to the Casimir effect. The Casimir
force which is due to it is \cite{Cas1}
$$
 F=-\frac{\pi^{2}}{240}\frac{1}{(2a)^{4}}. \eqno(86)
$$
Usually, the Casimir effect is validated experimentally for the
separation of parallel plates to be of order of $10^{-8}-10^{-5}
\rm m$, see, e.g., Ref.\cite{Bor}. Even if the mass of the
quantized charged matter field is estimated to be as small as the
electron mass (the Compton wavelength be of order of $10^{-12} \rm
m$), then the contribution of the latter to the Casimir effect is
damped as ${\rm e}^{-10^{4}} -\, {\rm e}^{-10^{7}}$, see (78), and
with stronger exponents, see (84). The contribution of (86) becomes
negligible at larger separations, where the contribution of the
quantized charged matter field dominates due to its independence
of the separation distance, see (76) or (82) (and (28) in
general). Although supercritical values of the magnetic field,
$|B|>m^{2}|e|^{-1}$, are hardly feasible in laboratory (but may be
attainable in some astrophysical objects such as magnetars), the
contribution of (76) may dominate for large enough but still
subcritical values of the magnetic field, $|B| \ll m^{2}|e|^{-1}$,
which can be attained in future in laboratory.

\section*{Acknowledgments}

We acknowledge the support from the National Academy of
Sciences of Ukraine (project No.0112U000054) and the State
Agency for Science, Innovations and Informatization of Ukraine
(SFFR-BRFFR grant F54.1/019). The work of Yu.~A.~S. was supported
by the ICTP -- SEENET-MTP grant PRJ-09
``Strings and Cosmology''.

\section*{Appendix. Abel-Plana summation formula}

We start by presenting the infinite sum over $l$ as an integral over
a contour in complex $w$-plane:
$$\sum\limits_{l=-\infty}^{\infty}f\left[\left(\pi{l}+\frac{\beta}{2}\right)^{2}\right]=\frac{1}{\pi}\int\limits_{C_{=}}{\rm d}w{f(w^{2})}G(w),\eqno(A.1)
$$where contour $C_{=}$ consists of two parallel infinite
lines going closely on the lower and upper sides of the real axis,
see Fig.1, and $G(w)$ possesses simple poles on the real axis at
$w=\pi{l}+\frac{\beta}{2}$:
$$G(w)=-\frac{1}{{\rm e}^{{\rm i}(2w-\beta)}-1}.\eqno(A.2)
$$By deforming the parts of contour $C_{=}$ on the $w$-plane into
contours $C_{\bigcap}$ and $C_{\bigcup}$ enclosing the lower and
upper imaginary semiaxes, see Fig.1, we get
$$\sum\limits_{l=-\infty}^{\infty}f\left[\left(\pi{l}+\frac{\beta}{2}\right)^{2}\right]=\frac{1}{\pi}\int\limits_{C_{\bigcap}}{\rm d}w{f(w^{2})}G(w)+\frac{1}{\pi}\int\limits_{C_{\bigcup}}{\rm d}w{f(w^{2})}G(w),\eqno(A.3)
$$where it is implied that all singularities of $f$ as a
function of $w$ lie on the imaginary axis at some distances from the
origin. In view of obvious relation
$$\lim\limits_{\epsilon\rightarrow{0+}}(\epsilon\pm{{\rm i}\nu})^{2}=\lim\limits_{\epsilon\rightarrow{0+}}(-\epsilon\mp{{\rm i}\nu})^{2}=(\pm{{\rm i}\nu})^{2}
$$for real positive $\nu$ and $\epsilon$, we obtain
$$\sum\limits_{l=-\infty}^{\infty}f\left[\left(\pi{l}+\frac{\beta}{2}\right)^{2}\right]=\frac{\rm i}{\pi}\int\limits_{0}^{\infty}{\rm d}\nu\{f[(-{\rm i}\nu)^{2}]-f[({\rm i}\nu)^{2}]\}[G(-{\rm i}\nu)-G({\rm i}\nu)].\eqno(A.4)
$$Let us note relation
$$G(-{\rm i}\nu)+G({\rm i}\nu)=1-g(\nu^{2}),\eqno(A.5)
$$where
$$G(\mp{{\rm i}\nu})=-\frac{1}{{\rm e}^{\pm{2\nu}-{\rm i}\beta}-1}\eqno(A.6)
$$and
$$g(\nu^{2})=\frac{{\rm i}\sin{\beta}}{\cosh{2\nu}-\cos{\beta}};\eqno(A.7)
$$note also that both $G(-{\rm i}\nu)$ and $g(\nu^{2})$ are exponentially
decreasing at large $\nu$. With the use of (A.5) we get
$$\sum\limits_{l=-\infty}^{\infty}f\left[\left(\pi{l}+\frac{\beta}{2}\right)^{2}\right]=\frac{\rm i}{\pi}\int\limits_{0}^{\infty}{\rm d}\nu\{f[(-{\rm i}\nu)^{2}]-f[({\rm i}\nu)^{2}]\}[2G(-{\rm i}\nu)+g(\nu^{2})]-$$
$$-\frac{\rm i}{\pi}\int\limits_{0}^{\infty}{\rm d}\nu{f[(-{\rm i}\nu)^{2}]}+\frac{\rm i}{\pi}\int\limits_{0}^{\infty}{\rm d}\nu{f[({\rm i}\nu)^{2}]}.\eqno(A.8)
$$
\begin{figure}
\includegraphics[width=390pt]{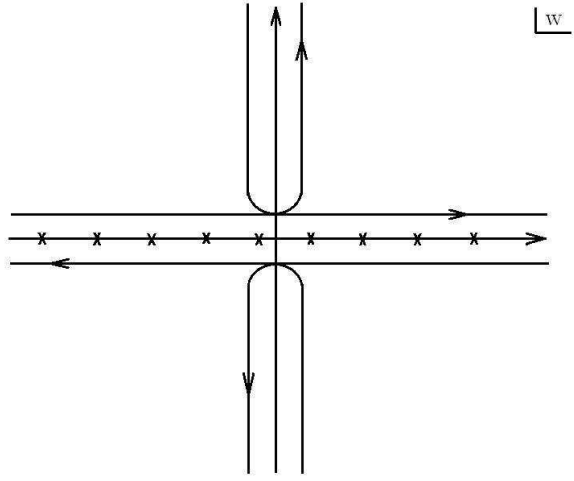}\\
\caption{Contours $C_{=}$, $C_{\bigcap}$ and $C_{\bigcup}$ on
the complex $w$-plane; the positions of poles of $G(w)$ are
indicated by crosses.}\label{1}
\end{figure}\normalsize
By rotating the paths of integration in the last and before the last
integrals in (A.8) by $90^{\circ}$ in the clockwise and
anticlockwise directions, respectively, we finally get
\newpage
$$\sum\limits_{l=-\infty}^{\infty}f\left[\left(\pi{l}+\frac{\beta}{2}\right)^{2}\right]=\frac{\rm i}{\pi}\int\limits_{0}^{\infty}{\rm d}\nu\{f[(-{\rm i}\nu)^{2}]-f[({\rm i}\nu)^{2}]\}[2G(-{\rm i}\nu)+g(\nu^{2})]+$$
$$+\frac{2}{\pi}\int\limits_{0}^{\infty}{\rm d}\mu{f(\mu^{2})}.\eqno(A.9)
$$In view of relation
$$2G(-{\rm i}\nu)+g(\nu^{2})=-\frac{\cos{\beta}-{\rm e}^{-2\nu}}{\cosh{2\nu}-\cos{\beta}},\eqno(A.10)
$$and the evenness of the integrand in the integral over $\mu$, we
obtain (56).

\end{document}